# Weighted Patterns as a Tool for Improving the Hopfield Model

Iakov Karandashev, Boris Kryzhanovsky and Leonid Litinskii

ya_rad_wsem@mail.ru, kryzhanov@mail.ru, litin@mail.ru

Scientific Research Institute for System Analysis Russian Academy of Sciences (Moscow)

### Abstract

We generalize the standard Hopfield model to the case when a weight is assigned to each input pattern. The weight can be interpreted as the frequency of the pattern occurrence at the input of the network. In the framework of the statistical physics approach we obtain the saddle-point equation allowing us to examine the memory of the network. In the case of unequal weights our model does not lead to the catastrophic destruction of the memory due to its overfilling (that is typical for the standard Hopfield model). The real memory consists only of the patterns with weights exceeding a critical value that is determined by the weights distribution. We obtain the algorithm allowing us to find this critical value for an arbitrary distribution of the weights, and analyze in detail some particular weights distributions. It is shown that the memory decreases as compared to the case of the standard Hopfield model. However, in our model the network can learn online without the catastrophic destruction of the memory.

## Introduction

The Hopfield model [1] is a well-known variant of an artificial associative memory. At the heart of the model is the Hebb connection matrix $\mathbf{J} = (J_{ij})$, which stores information about the set of $N$-dimensional input binary patterns:

$$J_{ij} = \frac{1-\delta_{ij}}{N} \sum_{\mu=1}^{M} x_i^\mu x_j^\mu, \quad x_i^\mu = \pm 1, \quad i, j = 1, 2, ..., N.$$

Here $\delta_{ij}$ is the Kronecker symbol, $\mathbf{x}^\mu = (x_1^\mu, x_2^\mu, ..., x_N^\mu)$ is the $\mu$-th input pattern and the number of patterns is equal to $M$. The input patterns are to be recognized by the network. In the second half of the 80th a storage capacity of the Hopfield model has been estimated by methods of statistical physics: the network is able to recognize about of $M_c = 0.138 \cdot N$ random patterns [2], [3]. Let us explain this result in more detail.

Let the number of patterns $M$ be less than the critical value $M_c$. If the neural dynamics starts with a binary configuration that is a distortion of a $k$-th pattern, the network rather quickly will be either in the pattern $\mathbf{x}^k$ itself, or in a configuration $\tilde{\mathbf{x}}^k$ that is very close to the pattern $\mathbf{x}^k$: $m(\mathbf{x}^k, \tilde{\mathbf{x}}^k) \approx 1$, where the overlap $m(\mathbf{x}^k, \tilde{\mathbf{x}}^k)$ is

$$m(\mathbf{x}^k, \tilde{\mathbf{x}}^k) = \frac{1}{N} \sum_{j=1}^{N} x_i^k \tilde{x}_i^k. \tag{1}$$

Theory shows [2]-[4] that when the number of patterns $M$ increases, at first the overlap $m(\mathbf{x}^k, \tilde{\mathbf{x}}^k)$ decreases slowly remaining of the order of 1. At the moment when the number of patterns $M$ exceeds a critical value $M_c$, the overlap $m(\mathbf{x}^k, \tilde{\mathbf{x}}^k)$ reduces abruptly almost to zero: $m(\mathbf{x}^k, \tilde{\mathbf{x}}^k) \approx 0$. Such phenomena are well-known in physics as phase transitions of the first kind (see [2]-[4]). The system in a jump-like way passes from the mixed phase (where all the patterns are metastable states) to the spin-glass phase (where all the local minima do not correlate with the patterns). This phase transition is regarded as a destruction of the memory. In the theory of neural networks it was named "catastrophic forgetting" [4]–[9].

The catastrophic forgetting is a troublesome defect of the Hopfield model. Indeed, let us imagine a robot whose memory is based on the Hopfield model. It is natural to assume that its memory is steadily filled up. When the robot sees a new image, it has to be written additionally to its memory. Catastrophic forgetting means that when the number of stored patterns exceeds $M_c$, the memory has to be completely destructed. Everything that was accumulated in the memory would be forgotten. This behavior contradicts the common sense.

Earlier some modifications of the Hebb matrix were proposed to eliminate the memory destruction [5]-[9]. These models allow one to write down one after another an unlimited number of random patterns into the matrix elements. However, the real memory of the network is restricted. If as previously the maximum number of recognized patterns is denoted by $M_c$, for the models discussed in [5]-[9] the value of $M_c \approx 0.05 \cdot N$. All these models have the same weak point: only those patterns that are the last written down in the connection matrix constitute the real memory of the network. Let us explain the last statement. Suppose the patterns are enumerated in order of their appearance during the learning process: the later the pattern is shown to the network, the larger is its number. Then it turns out that the real network memory is formed only of the patterns whose numbers are in the interval $\mu \in [M - M_c, M]$.

Patterns with the order numbers less than $M - M_c$ are irretrievably excluded from the memory. That is the common property of the models [5]-[9].

In other words, such a memory has the form of a moving window whose width is about $0.05 \cdot N$. The first shown patterns are removed from the memory. Can these patterns be restored? What happens if they are again shown to the network? From general reasons it follows that the anew shown patterns cannot be recognized. However, up to now we do not have the irrefragable answer to this question.

In our work we succeeded in eliminating the catastrophic forgetting typical for the Hopfield model. We supply every pattern by an individual weight that is proportional to the frequency of the pattern appearance at the input of the network. Then in place of the Hebb matrix we obtain a quasi-Hebbian matrix of the form

$$J_{ij} = (1-\delta_{ij})\sum_{\mu=1}^{M} r_\mu x_i^\mu x_j^\mu . \qquad (2)$$

The weights $r_\mu$ are positive and put in decreasing order: $r_1 \geq r_2 \geq ... \geq r_M \geq 0$. In computer simulation it is convenient to use the normalization condition $\sum_{\mu=1}^{M} r_\mu^2 = 1$.

The main result is the following: formula (2) allows one to write down any number of patterns in a matrix without the catastrophic forgetting. At the same time, only those patterns whose weights exceed a critical value $r_c$ would be recognized. The critical weight $r_c$ depends on the distribution of the weights.

This approach allows the network to learn continuously, even during the process of pattern recognition. Each pattern which appears at the input of the network is automatically added to the matrix. The distribution of weights is changing continually, but no catastrophic destruction of the memory occurs. At any time the real memory of the network consists of the patterns that have been shown to the network quite often.

In Section II with the aid of statistical physics methods the main equation for the quasi-Hebbian connection matrix (2) is derived. To verify our equation we examine again the well-known case of the standard Hopfield model ($r_\mu \equiv 1$). In Section III we analyze the case when only one weight is different from the others, which are identical $r_1 \neq r_2 = r_3 = ... = r_M = 1$. Here the complete analytical analyses can be performed. We show that the pattern with the unique weight substantially affects the properties of the network memory. Computer simulations confirm the theoretical results. In Section IV we analyze the main equation for an arbitrary weights distribution. The main result is as follows: For every distribution of the weights there is such a critical value $r_c$ that only patterns whose weights are greater than $r_c$ would be recognized by the network. Other patterns are not recognized. We succeeded to find only the algorithm of calculation of the critical value $r_c$. It is not clear whether it is possible to obtain an analytical expression for $r_c$ in the general case. Some particular distributions of the weights are discussed in Section V in detail.

Note, for the first time the quasi-Hebbian connection matrix (2) was discussed many years ago. For this matrix the implicit form of the main equation (5) was obtained in [8]. However, in [8] the authors examined the case of the standard Hopfield model only ($r_\mu \equiv 1$). Our contribution is the solution of the main equation in the general form.

## II. Main Equation and the Standard Hopfield Model

**1. The main equation.** Let $\mathbf{S} = (S_1, S_2, ..., S_N)$, where $S_i = \pm 1$, define a state of the system as a whole. The energy of the state can be presented in the form

$$E(\mathbf{S}) = -\frac{1}{2}\sum_{\mu=1}^{M} r_\mu \left( m_\mu^2(\mathbf{S}) - \frac{1}{N} \right), \qquad (3)$$

where $m_\mu(\mathbf{S})$ is the defined in Eq.(1) overlap of the current state $\mathbf{S}$ with the input pattern $\mathbf{x}^\mu$:

$$m_\mu(\mathbf{S}) = \frac{1}{N}\sum_{i=1}^{N} S_i x_i^\mu .$$

Fixed points of the network are the local minima of the energy (3). Statistical physics methods allow one to obtain equations for the overlap of a local minimum with the patterns $\{\mathbf{x}^\mu\}_1^M$ (see [2]-[4]). After solving these equations it is possible to understand under which conditions the overlap of the local minimum with the $k$-th pattern is of the order of 1 ($m_k \sim 1$). In other words: under which conditions the local minimum coincides (or nearly coincides) with the $k$-th input pattern.

Suppose the dimensionality of the problem is very large: $N \gg 1$. After repeating step by step the calculations performed in [2]-[4] for the standard Hopfield model, we obtain the system of equations for the $k$-th pattern (in the zero-temperature limit $T \to 0$):

$$m_k = \text{erf}\left(\frac{r_k m_k}{\sqrt{2}\sigma_k}\right),$$

$$\sigma_k^2 = \frac{1}{N}\sum_{\mu \neq k}^{M} \frac{r_\mu^2}{(1-C \cdot r_\mu)^2}, \tag{4}$$

$$C = \frac{1}{\sigma_k}\sqrt{\frac{2}{\pi}} \exp\left[-\left(\frac{r_k m_k}{\sqrt{2}\sigma_k}\right)^2\right].$$

Here $m_k$ is the overlap of the local minimum with the $k$-th pattern, $\sigma_k^2 = \sum_{\mu \neq k}^{M} r_\mu^2 m_\mu^2$ is the weighted sum of squared overlaps of all patterns except the $k$-th one, and $C = \lim_{T \to 0}(1-q)/T$, where $q$ is the Edwards-Anderson order parameter [10]. Equations (4) are obtained in the replica symmetry approximation. In the case of equal weights ($r_\mu \equiv 1$) this system reduces to the well-known system for the standard Hopfield model (see Eqs. (2.71)-(2.73) in [4]).

Let us introduce an auxiliary variable $y = r_k m_k / \sqrt{2}\sigma_k$. Excluding $\sigma_k$ and $C$ from the system (4), we obtain the equation

$$\frac{1}{\alpha} = \frac{1}{\gamma^2} \cdot \frac{1}{M-1}\sum_{\mu \neq k}^{M} \frac{r_\mu^2}{(r_k \cdot \varphi - r_\mu)^2}. \tag{5}$$

As usual by $\alpha = M/N$ we define the load parameter; $\gamma = \gamma(y)$ and $\varphi = \varphi(y)$ are equal to:

$$\gamma(y) = \sqrt{\frac{2}{\pi}} e^{-y^2}, \quad \varphi(y) = \frac{\sqrt{\pi}}{2} \frac{\text{erf } y}{y} e^{y^2}. \tag{6}$$

The function $\gamma = \gamma(y)$ decreases monotonically, and the function $\varphi = \varphi(y)$ increases monotonically beginning from its minimal value $\varphi(0) = 1$. For simplicity sometimes we omit the argument of these functions.

Let us fix the values of the external parameters: $N$, $\alpha = M/N$, $\{r_\mu\}_1^M$ and $k$. If $y_0$ is a solution of Eq. (5), the overlap of the local minimum with the $k$-th pattern is equal to

$$m_k = \text{erf}(y_0). \tag{7}$$

When $y_0$ is known, we can calculate the value of $\sigma_k = r_k \text{erf } y_0 / \sqrt{2} y_0$ entering into Eq.(4).

It is important to determine the critical value of the load parameter $\alpha_c$ for which the solution of Eq. (5) still exists, however if $\alpha$ is larger than $\alpha_c$ there is no solution of Eq.(5). All characteristics corresponding to the critical value $\alpha_c$ are indicated by the same subscript "c". They are $y_c$, $m_c$ and $\sigma_c$.

**2. The standard Hopfield model:** $r_\mu \equiv 1$. Since all patterns are equivalent, the subscript $k$ in Eqs. (4), (5) can be omitted. It is easy to see that in this case Eq. (5) has the form

$$\alpha = \gamma^2 (\varphi - 1)^2 \tag{8}$$

The plot of the rhs of Eq. (8) is shown in Fig. 1. We see that for all $\alpha$ (that are less than the critical value) there are two solutions: $\bar{y}_\alpha$ and $y_\alpha$ ($\bar{y}_\alpha < y_\alpha$). We are interested in the larger solution $y_\alpha$ only. The solution $\bar{y}_\alpha$ is a spurious one, and it has to be omitted.

When $\alpha$ increases the solution $y_\alpha$ shifts to the left. Consequently, the overlap $m_k$ (7) decreases. In other words, the local minimum little by little moves away from the pattern. Increasing $\alpha$, we reach the critical value $\alpha_c$, which is equal to the maximal value of the function in the rhs of Eq. (8). It is not difficult to calculate the critical value: $\alpha_c \approx 0.138$. This well-known result has been obtained for the first time in [2], [3].

What happens when the load parameter $\alpha$ exceeds the critical value $\alpha_c$? In this case there is no solution of Eq. (8). As they say, "a breakdown" of the solution happens. The detailed analysis shows ([2]-[4]) that the overlap abruptly decreases almost up to zero. Local minima do not correlate with the input patterns anymore. In other words, the memory of the network is completely destructed.

The expression in the rhs of Eq. (8) reaches its maximum value at the point $y_c \approx 1.511$. Equation (7) allows one to calculate the critical value of the overlap: $m_c \approx 0.967$. We see that even in the worst case of the maximal value of the load parameter $\alpha_c$, the overlap of a pattern with the nearest local minimum is large enough. If the value of the load parameter is less than $\alpha_c$, the overlap is even closer to 1. Since we considered an arbitrary pattern, we can say that while $\alpha < \alpha_c$ there is a local minimum in the vicinity of each pattern. It can be shown that the depths of these local minima are approximately the same [4].

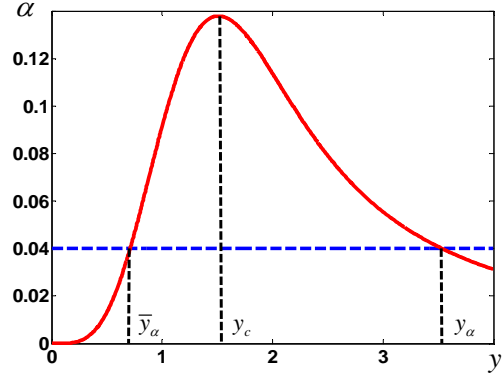

**Fig. 1.** Graphical solution of Eq. (8) for the standard Hopfield model.
Solid line is the graph of the rhs of Eq.(8). Dashed horizontal straight line corresponds to $\alpha = 0.04$.

**Note.** Of the two solutions of Eq. (8) we use only one that is $y_\alpha$. The rejected one, $\bar{y}_\alpha$, is the spurious solution: when the load parameter $\alpha$ decreases, it tends to 0, and, consequently, has no physical meaning. A more general statement is as follows. Equation (5) always has several solutions, and all of them correspond to the stationary points of the saddle-point equation [4]. However, if there are more than two solutions, only the one lying to the right of the rightmost maximum of the rhs of Eq. (5) corresponds to the minimum of the free energy. We use this in the next Section.

## III. One Pattern with Unique Weight

Interesting is the case when all the weights $r_\mu$, except one, are equal to each other, and only one weight differs from the others. Without loss of generality we can write

$$r_1 = \tau, \quad r_2 = r_3 = ... = r_M = 1. \tag{9}$$

The unique weight $\tau$ can be both larger and less than 1.

It might appear that in this case the difference with the standard Hopfield model has to be negligibly small: an enormous number of patterns with the same weights enter the expression for the connection matrix and only one pattern provides a different contribution. However, this is not the case. One pattern with a unique weight $\tau$ can substantially change a network memory. Let us examine separately what happens to the local minimum in the vicinity of the pattern with the unique weight $\tau$, and what happens to the local minima near other patterns with the same weights.

**1. Pattern with a unique weight:** $r_1 = \tau$. For this pattern equation (5) has the form

$$\alpha = \gamma^2 (\tau \cdot \varphi - 1)^2. \tag{10}$$

If $y$ is a solution of Eq. (10), the overlap of the local minimum with the first pattern is $m^{(1)} = erf(y)$. The superscript "(1)" emphasizes that we deal with the overlap with the pattern number 1.

The point of the breakdown $y_c^{(1)}(\tau)$, the one where the rhs of Eq.(10) reaches its maximum, is the solution of the equation

$$\varphi(y) = 1 + \frac{2y^2}{\tau}. \tag{11}$$

After finding $y_c^{(1)}(\tau)$, the critical characteristics $m_c^{(1)} = erf(y_c^{(1)})$ and $\alpha_c^{(1)}$ can be calculated.

In Fig.2a we present graphs of the rhs of Eq. (10) for different values of the weight $\tau > 1$. We see that when $\tau$ increases, the critical value $\alpha_c^{(1)}(\tau)$ increases too. In the same time the breakpoint $y_c^{(1)}(\tau)$ steadily moves toward zero. It turns out that Eq.(11) has a nontrivial solution only if $\tau < 3$.

Suppose, for example, that $\tau = 2$. Then the breakpoint $y_c^{(1)}(2) \approx 1$ and the critical characteristics are equal to $\alpha_c^{(1)} \approx 0.805$ and $m_c^{(1)} \approx 0.84$. As long as the load parameter $\alpha$ does not exceed the critical value, $\alpha \leq 0.805$, Eq. (10) has a solution in the region $y \geq 1$, and the overlap is relatively large (it is larger than $m_c^{(1)}$). In other words, the first pattern is recognized. If the load parameter is larger than the critical value, $\alpha > 0.805$, Eq.(10) has no solutions, and the first pattern is not recognized. The breakdown of the solution occurs because when $\alpha$ crosses the value $\alpha_c^{(1)}$ the system undergoes the phase transition of the first kind. This is true for each $\tau < 3$.

Let us examine now what happens when $\tau \geq 3$. For such $\tau$ Eq.(11) has no solutions at all and the phase transition in the system vanishes. In the same time for a fixed value $\tau \geq 3$, Eq.(10) has a solution when $\alpha$ is less than its limiting value that coincides with the value of the maximum of the rhs of Eq.(10). For $\tau \geq 3$ this maximum is always in the point $y = 0$. Again, by $\alpha_c^{(1)}(\tau)$ we denote the limiting value of $\alpha$. From the expression in the rhs of Eq.(10) it follows that for $\tau \geq 3$ we have $\alpha_c^{(1)}(\tau) = 2(\tau-1)^2/\pi$. So, for a fixed value $\tau \geq 3$ the overlap of the first pattern does not equal to zero only if $\alpha$ is less than the limiting value $2(\tau-1)^2/\pi$. If, on the contrary, $\alpha \geq 2(\tau-1)^2/\pi$, the overlap is equal to zero and the pattern is not recognized.

Now let us examine the weights $\tau$ from the interval $[0,1]$. In Fig.2b we present graphs of the rhs of Eq. (10) for different weights $\tau \leq 1$. We see, that when $\tau$ decreases from 1 to 0, the point of the rightmost maximum $y_c^{(1)}(\tau)$ steadily shifts to the right. Consequently, the critical value of the overlap $m_c^{(1)}(\tau)$ tends to 1. In the same time the critical value of the load parameter $\alpha_c^{(1)}(\tau)$ steadily decreases. Note we are not interested in the behavior of the curve to the left from the rightmost maximum, since in this region Eq. (10) has only spurious solutions.

So, when $\tau$ increases from 0 to 3, the critical value of the overlap $m_c^{(1)}(\tau)$ decreases. Let us point out that this is true only for the critical values $y_c^{(1)}(\tau)$ and $m_c^{(1)}(\tau)$, i.e. to the value of the overlap at the breakdown point $\alpha = \alpha_c^{(1)}(\tau)$. Absolutely other situation arises when $\alpha$ is less than the critical value: $\alpha < \alpha_c^{(1)}(\tau)$. Indeed, for example, in Fig.2a we see intersections of a straight line $\alpha = 0.5$ and the graphs representing the rhs of Eq. (10) for different values of $\tau > 1$. We see that when $\tau$ increases the point of intersection shifts to the right. This means that when $\tau$ increases the overlap $m^{(1)}(\tau)$ also increases. This behavior of the overlap is in agreement with the common sense: the greater the weight of the pattern $\tau$, the greater its influence. Then the overlap of this pattern with the local minimum has to be greater.

Let us summarize the obtained results. In contrast to the standard Hopfield model we have two independent free parameters. They are the load parameter $\alpha$ and the weight $\tau$. Let us fix the value of $\tau$ and change the load parameter $\alpha$. If $\tau < 3$, then for some value of $\alpha_c^{(1)}(\tau)$ the solution of the set of equations (10), (11) vanishes. This breakdown of the solution means that the system undergoes a phase transition of the first kind. In this case the pattern is ceased to be recognized. However, if the fixed value of $\tau$ is larger than 3, $\tau \geq 3$, Eq.(11) has no solutions. When $\alpha$ increases, the value of the overlap decreases smoothly, and it becomes equal to zero beginning from the value $\alpha_c^{(1)}(\tau) = 2(\tau-1)^2/\pi$. There is no phase transition of the first kind when $\tau \geq 3$.

Let us on the contrary fix $\alpha$ and vary $\tau > 0$. From Fig.2a, b it follows that if we steadily increase $\tau$, sooner or later we reach the value $\tau(\alpha)$ beginning of which a solution of Eq.(10) appears. For $\alpha < 8/\pi$ the point of contact is equal to the critical value $y_c^{(1)} > 0$. For example, in Fig.2a we fix $\alpha = 0.5$. Then $y_c^{(1)}$ is the point of contact of the straight line $\alpha = 0.5$ with the curve that is the rhs of Eq.(10) for $\tau = 1.66$. In the point of contact the overlap increases in a jump-like way from zero to $m_c^{(1)} = erf(y_c^{(1)})$ (also see the jumps of the overlaps in Fig.5). However, if the fixed value of $\alpha$ exceeds $8/\pi$, the contact of the line $\alpha = const$ with the abovementioned curve occurs in the point $y = 0$ and the corresponding overlap is equal to zero: $m^{(1)} = 0$. The following increase of the weight $\tau$ leads to the steady increase of the overlap $m^{(1)}$ up to its maximal value equal to 1.

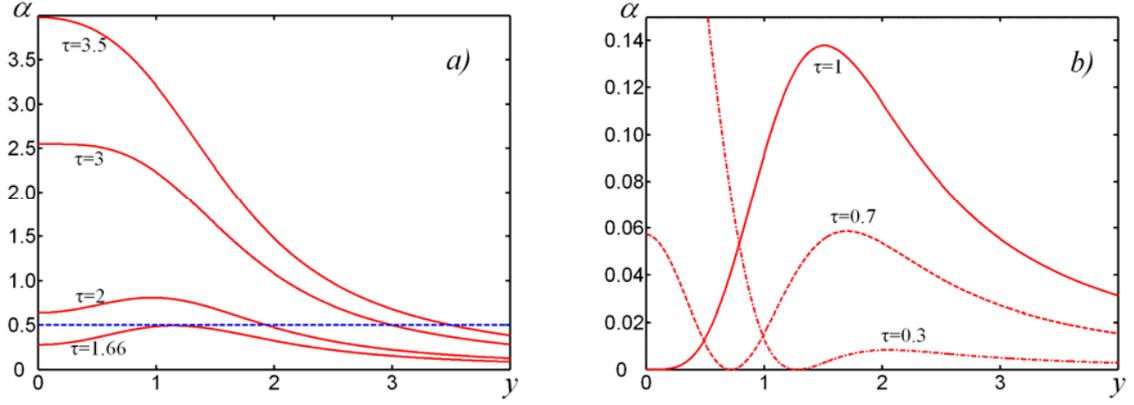

**Fig. 2.** The graphs of the right-hand side of Eq. (10) for different values of $\tau$: *a)* $\tau \geq 1$; the straight line intersects the graphs corresponding to different values of $\tau$; *b)* $\tau \leq 1$.

In two left panels of Fig.3 we show the critical characteristics $\alpha_c^{(1)}(\tau)$ and $m_c^{(1)}(\tau)$ as functions of $\tau$. Let us emphasize that at the point $\tau = 3$ in Fig.3a two different curves conjugate smoothly. To the left of $\tau = 3$ is the curve that corresponds to the critical values of the load parameter $\alpha_c^{(1)}(\tau)$, for which the breakdown of the solution occurs. These values of $\alpha_c^{(1)}(\tau)$ are obtained as a result of numerical solution of the system of equations (10)-(11). To the right of the point $\tau = 3$ is the curve $\alpha_c^{(1)}(\tau) = 2(\tau-1)^2/\pi$.

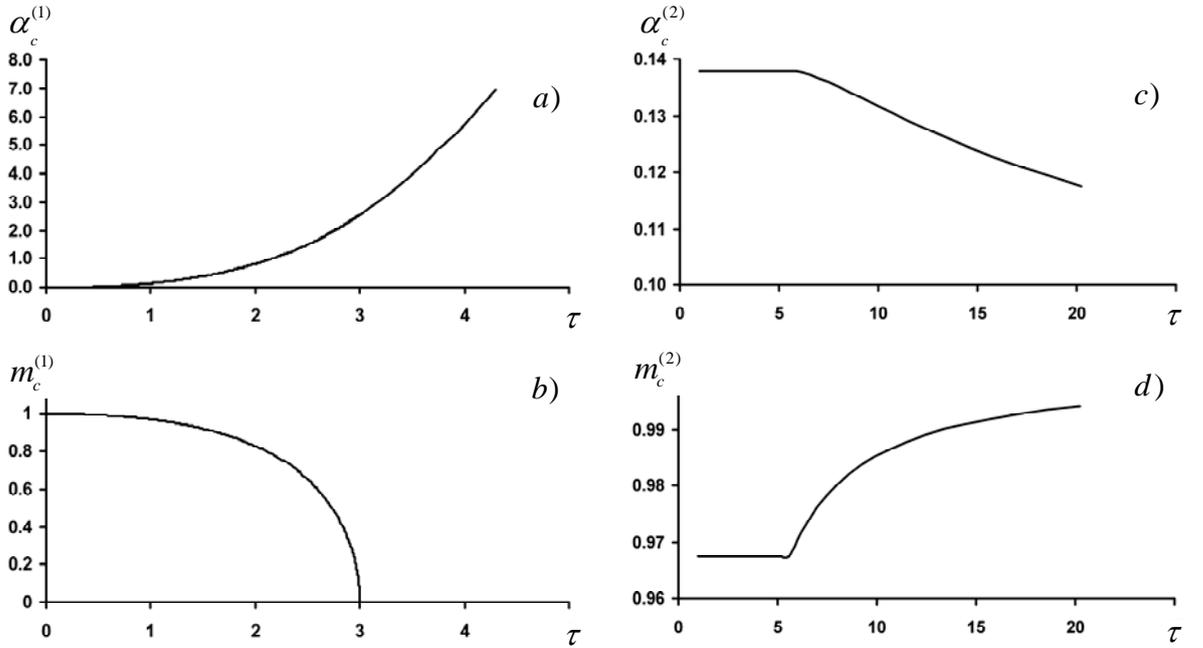

**Fig. 3.** Critical values of the load parameter $\alpha_c(\tau)$ (two upper panels) and overlap $m_c(\tau)$ of the local minimum with the pattern (two lower panels). The curves on the left panels (*a* and *b*) correspond to the pattern with a unique weight $r_1 = \tau$; the curves on the right panels (*c* and *d*) correspond to the patterns with the same weights $r_\mu = 1$, $\mu \geq 2$.

**2. Patterns with the same weights:** $r_\mu = 1$, $\mu \geq 2$. Let us examine how the overlap of the local minimum with one of the patterns whose weight is equal to 1 depends on the value of $\tau$. Since all these patterns are equivalent, we choose the pattern with number "2". The superscript (2) indicates the characteristics $m^{(2)}, y^{(2)}, \alpha^{(2)}$ that are interesting for us. Now Eq. (5) has the form:

$$\alpha = L(y) \equiv \frac{\gamma^2 (\varphi-1)^2 (\varphi-\tau)^2}{(1-\varepsilon)(\varphi-\tau)^2 + \varepsilon\tau^2 (\varphi-1)^2}, \tag{12}$$

where $\varepsilon = 1/M$. When $M \to \infty$, the value of $\varepsilon$ tends to zero. However, we cannot simply put $\varepsilon = 0$, since then for some value of $y$ the denominator of $L(y)$ necessarily vanishes (at least when $\tau > 1$), and then it is impossible to cancel $(\varphi-\tau)^2$ in the numerator and the denominator of $L(y)$. So, we analyze Eq. (12) for a small, but finite value of $\varepsilon$ and after this we let it tend to zero. This way of analysis is correct.

When $\tau \leq 1$ the function $\varphi(y) - \tau$ has no zeros, and we can simply take $\varepsilon = 0$. In this case the expression for $L(y)$ turns into $\gamma^2(\varphi-1)^2$. Then Eq. (12) transforms into Eq. (8), and the last corresponds to the standard Hopfield model. Consequently, until $\tau \in (0,1]$ the solution of Eq.(12) does not depend on $\tau$ at all, and it coincides with the solution for the standard Hopfield model. The critical characteristics do not depend on $\tau$ as well and they are equal:

$$y_c^{(2)}(\tau) \equiv y_c \approx 1.511, \ \alpha_c^{(2)}(\tau) \equiv \alpha_c \approx 0.138, \ m_c^{(2)}(\tau) \equiv m_c \approx 0.967. \tag{13}$$

Let us examine the interval $\tau > 1$. In this case the function $\varphi(y) - \tau$ vanishes at the point $y_0(\tau)$. The value of the argument $y_0(\tau)$ is determined by equation:

$$\varphi(y_0(\tau)) = \tau \ \Leftrightarrow \ y_0(\tau) = \varphi^{-1}(\tau), \tag{14}$$

where $\varphi^{-1}$ is the inverse of $\varphi$. Out of the small vicinity of the point $y_0(\tau)$ the parameter $\varepsilon$ in Eq. (12) can be allowed to tend to zero with the function $L(y)$ being the same as in the case of the standard Hopfield model: $L(y) = \gamma^2(\varphi-1)^2$. On the other hand, for small, but nonzero value of $\varepsilon$ at the point $y_0(\tau)$ the function $L(y)$ itself is equal to zero: $L(y_0(\tau)) = 0$. If $y$ is from the vicinity of the point $y_0(\tau)$ and it tends to $y_0(\tau)$, for any nonzero value of $\varepsilon$ the curve $L(y)$ quickly drops to zero. Thus, for any nonzero value of $\varepsilon$ the graph of the function $L(y)$ practically everywhere coincides with the "standard" curve $\gamma^2(\varphi-1)^2$, but in the vicinity of the point $y_0(\tau)$ the curve $L(y)$ has a narrow dip down to zero (whose width is proportional to the value of $\varepsilon$).

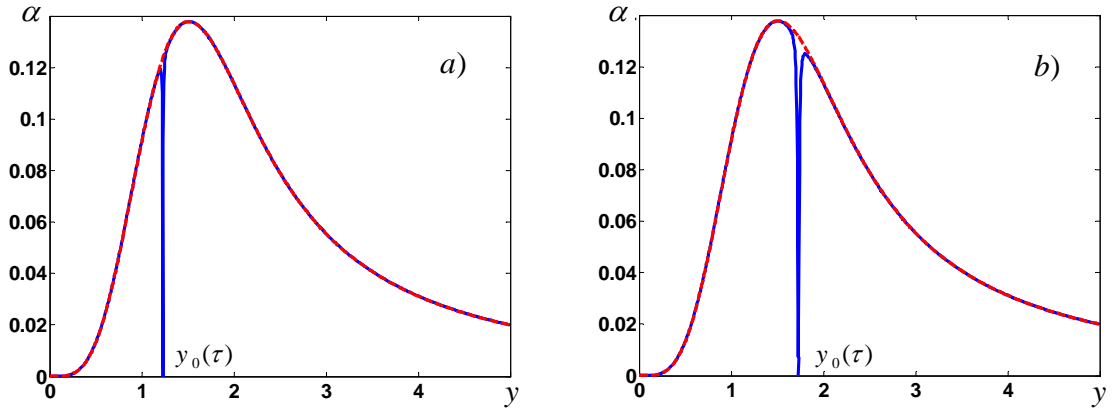

**Fig. 4.** The graph of the function $L(y)$ from Eq. (12) for $\varepsilon = 10^{-5}$ (solid line): a) when $\tau = 3 < \varphi(y_c)$ the point $y_0(\tau)$ is on the left of $y_c \approx 1.511$; b) when $\tau = 10 > \varphi(y_c)$ the point $y_0(\tau)$ is on the right of $y_c$. Dashed line shows the difference with the standard Hopfield model.

As long as the weight $\tau < \varphi(y_c) \approx 5.568$, the point where the function $\varphi(y) - \tau$ is equal to 0 is to the left of $y_c = 1.511$: $y_0(\tau) < y_c$. For this case in Fig. 4a we show the curve $L(y)$. The rightmost maximum of the curve $L(y)$ corresponds to the critical point $y_c \approx 1.511$ and it does not depend on $\tau$. Consequently, the equalities (13) are valid not only for $\tau \in [0,1]$, but in the wider interval $0 < \tau \leq \varphi(y_c) \approx 5.568$.

On the contrary, for the values of the weight $\tau > 5.568$ the point $y_0(\tau)$ is to the right of $y_c$. For this case an example of the curve $L(y)$ is shown in Fig. 4b. The rightmost maximum of the curve $L(y)$ coincides with the peak of the curve that is slightly to the right of the point $y_0(\tau)$. From continuity conditions it is evident that when $\varepsilon \to 0$ this peak shifts to the point $y_0(\tau)$. Consequently, when $\varepsilon \to 0$ for $\tau > 5.568$ we have

$$y_c^{(2)}(\tau) \equiv y_0(\tau) > 1.511, \ m_c^{(2)}(\tau) \equiv erf(y_c^{(2)}(\tau)) > 0.967, \ \alpha_c^{(2)}(\tau) = \frac{2}{\pi}(\tau-1)^2 e^{-2y_0^2(\tau)} < \alpha_c = 0.138. \tag{15}$$

Let us summarize the results obtained for patterns with the same weights. Firstly, as far as the value of $\tau$ is less than $\varphi(y_c) = 5.568$, all characteristics of the local minima do not depend on $\tau$ and coincide with the characteristics of the standard Hopfield model. Secondly, as soon as the value of $\tau$ exceeds $\varphi(y_c) = 5.568$ the situation changes. In this case the breakdown point of the solution depends on $\tau$ and coincides with $y_0(\tau)$ that is the solution of Eq. (14). When $\tau$ increases the critical value of the load parameter $\alpha_c(\tau)$ decreases and the overlap with the pattern increases and tends to 1 (see Eq. (15)). The graphs showing the behavior of the critical characteristics $m_c^{(2)}(\tau)$ and $\alpha_c^{(2)}(\tau)$ are presented at the two right panels in Fig.3.

**3. Computer simulations.** The obtained results were verified with the aid of computer simulations. For a given value of $N$ the load parameter $\alpha$ was fixed. Then $M = \alpha N$ randomized patterns were generated, and they were used to construct the connection matrix with the aid of Eqs. (2) and (9). When choosing the weight coefficient $\tau$ we proceeded from the following reasons.

Let $\alpha$ be a fixed value of the load parameter. Then we found the value of the weight $\tau(\alpha)$ for which the given $\alpha$ was a critical one. This can be done by solving Eqs. (10) and (11) simultaneously. We also defined the breakdown point $y_c(\alpha)$ of the solution. (For example, Fig.2a shows that the value of the load parameter $\alpha = 0.5$ is a critical one for the weight $\tau(\alpha) = 1.66$, while the breakdown point is equal to $y_c(\alpha) \approx 1.15$.) When constructing the connection matrix with the weight $\tau$ that is less than $\tau(\alpha)$, the mean value of the overlap with the pattern has to be close to 0. If the weight $\tau$ is equal to $\tau(\alpha)$, the mean value of the overlap has to be close to $m_c(\alpha) = erf(y_c(\alpha))$. If the weight $\tau$ is larger than $\tau(\alpha)$, the mean value of the overlap has to be larger than $m_c(\alpha)$. Under further increase of $\tau$ the mean overlap has to tend to 1. In other words, the main idea of the computer simulations is as follows: For a fixed value of $\alpha$ we vary the weight $\tau$ near the critical value $\tau(\alpha)$ to verify that our numerical results are close to theoretical predictions.

**3.1. Pattern with unique weight.** To verify the theoretical conclusions for the pattern with the unique weight $r_1 = \tau$, three experiments have been done for three different values of $\alpha$. For each $\alpha$ we varied $\tau$ in the region of the critical value $\tau(\alpha)$ and with the aid of computer simulations examined how the mean value of the overlap of the pattern with the nearest local minimum $<m>$ depends on $\tau$. For this purpose for each fixed value of $\tau$ we generated a set of random matrices. For each matrix we started the neural dynamics from the pattern with unique weight $\tau$ and found the local minimum, and then we calculated its overlap with the pattern. The mean value of the overlap $<m>$ was calculated by averaging over the set of all random matrices. We did our simulations for three different dimensionalities: $N = 1000$, $10000$ and $30000$. For $N = 1000$ we averaged over 100 matrices, for $N = 10000$ over 20 matrices and for $N = 30000$ over 10 matrices (the complexity of calculations increases as $O(N^3)$, and this is why we had to decrease the number of random matrices).

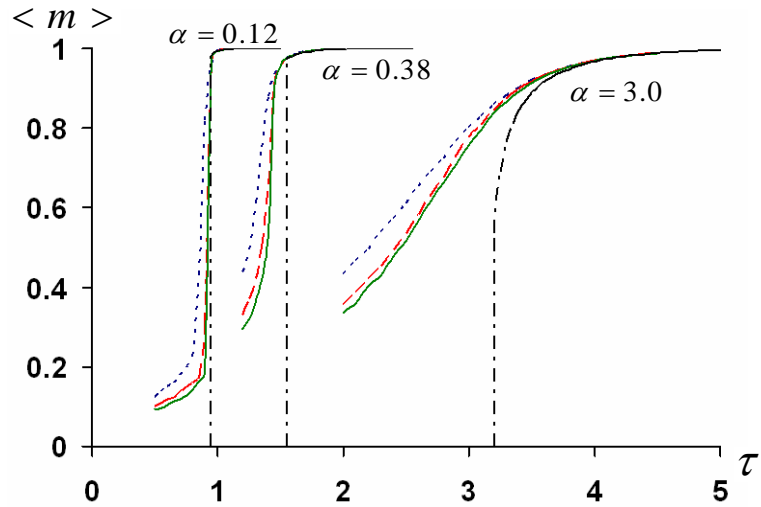

**Fig. 5.** Theoretical and experimental results for the pattern with the unique weight $r_1 = \tau$. The graphs correspond to three different values of the load parameter $\alpha$ and to three different dimensionalities $N$. Chained lines correspond to theoretical results. For experimental results: dotted lines correspond to $N = 1000$, dashed lines correspond to $N = 10000$, and solid lines to $N = 30000$.

Let us list the values of $\alpha$ that were chosen for testing with the theoretical values of $\tau(\alpha)$ and $m_c(\alpha)$: 1) $\alpha = 0.12$, $\tau(\alpha) \approx 0.944$, $m_c^{(1)}(\alpha) \approx 0.971$; 2) $\alpha = 0.38$, $\tau(\alpha) \approx 1.501$, $m_c^{(1)}(\alpha) \approx 0.919$; 3) $\alpha = 3.0$; for this value of $\alpha$ there is no jump of the overlap, but beginning from $\tau(\alpha) \approx 3.171$, the overlap has to increase smoothly.

In Fig.5 the graphs for all three values of $\alpha$ are presented. Theoretical curves $m^{(1)}(\tau)$ are shown by chained lines. The results of our computer simulations are given by dotted lines for $N = 1000$, by dashed lines for $N = 10000$ and by solid lines for $N = 30000$.

For $\alpha = 0.12$ and $\alpha = 0.38$ on the experimental curves we clearly see the expected jumps of the mean overlaps that have place in the vicinities of the critical values of the weights $\tau(\alpha) \approx 0.944$ and $\tau(\alpha) \approx 1.501$, respectively. To the left from the jump-point the values of the mean overlap are not equal to zero. Firstly, this can be due to not sufficiently large dimensionality of the problem. (The point is that all theoretical results are valid in the thermodynamic limit $N \to \infty$. For computer simulations we used very large, but finite dimensionality $N$. Note when $N$ increases the experimental curve is becoming more aligned with the theoretical step function.) Secondly, the theory is correct when the mean overlap is close to 1 (but not to 0). For these values of $\tau$ we have rather good agreement between the theory and the computer simulations.

For the third value of the load parameter $\alpha = 3.0$ we expected a smooth increase of the mean overlap $<m>$ from 0 to 1. Indeed, all the rightmost experimental curves increase smoothly without any jumps. However, according to our theory the increase ought to start beginning from $\tau \approx 3.1$, while the experimental curve starts to deviate from zero much earlier. This discrepancy between our theory and the experiment can be explained in the same way as it has been done in the end of the previous paragraph. When the dimensionality $N$ increases, the experimental curve again approximates the theoretical curve better.

**3.2. Patterns with the same weights.** To verify our theory in the case of patterns with the same weights ($r_\mu = 1, \mu \geq 2$) we used an analogous procedure. For the value of the load parameter $\alpha = 0.12$ and several dimensionalities $N$ ($N = 3000$, 10000 and 30000) we calculated the mean overlap $<m>$ between the pattern and the nearest local minimum for different values of $\tau$. We averaged both over $M-1$ patterns of the given matrix and over 10 randomized matrices constructed for the given value of $\tau$.

According to the theory, for $\alpha = 0.12$ the breakdown of the overlap $<m>$ has to take place when $\tau \approx 17.1$. In Fig. 6 this is shown by the right dashed straight line with the label "$M \to \infty$". If $N$ and $M$ are indeed infinitely large, just in this place the breakdown of $<m>$ has to take place. The dependency of $<m>$ on $\tau$ observed in our experiments is shown in Fig.6 by three solid lines corresponding to different dimensionalities $N$.

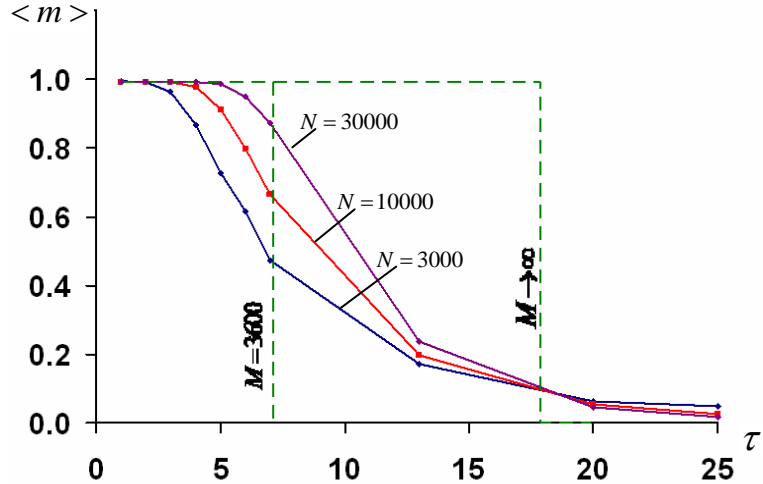

**Fig. 6.** Theory and experiments for patterns with the same weights $r_\mu = 1$ for $\alpha = 0.12$. Solid lines show the results of the experiments for three values of dimensionality $N$.

This noticeable difference between the theory and computer simulations must not confuse us. Apparently this difference is due to finite dimensionalities of the experimental connection matrices. Earlier computer verifications of classical theoretic results faced just the same problems [11], [12]. As a way out the authors of these papers extrapolated their experimental results into the region of very large dimensionalities $N$.

Note, when $N$ increases the experimental curve in Fig. 6 tends to "theoretical step-function", which is indicated with the aid of dashed line. A correction due to a finite dimensionality of the problem can be taken into account if we insert the explicit expression $\varepsilon = 1/M$ in Eq. (12). Then for $N = 30000$ we have $M = \alpha N = 3600$. When this value of $M$ is used in Eq. (12), we obtain that the breakdown of the overlap $<m>$ has to take place not in the vicinity of $\tau \approx 17.1$, but much earlier, when $\tau \approx 7.1$. The corresponding dashed line with the label "$M = 3600$" is shown in Fig. 6. Its location noticeably better correlates with the experimental curves.

## IV. Arbitrary Distribution of Weights

In this Section we present the method of solving of Eq. (5) in the general case. We show that there is a critical weight $r_c$, so that the memory of the network consists only of the patterns whose weights exceed $r_c$. Patterns whose weights are less than $r_c$, are not recognized by the network.

Let us transform the Eq.(5) dividing the left-hand and the right-hand sides by $M$. Then the main equation takes the form:

$$N = \sum_{\mu \neq k}^{M} f_\mu^{(k)}(y), \qquad (16)$$

where $f_\mu^{(k)}$ are functions of $\gamma$, $\varphi$ and $r_\mu$:

$$f_\mu^{(k)}(y) = \left( \frac{t_\mu^{(k)}}{\gamma(\varphi - t_\mu^{(k)})} \right)^2, \quad t_\mu^{(k)} = \frac{r_\mu}{r_k}, \quad \mu \neq k. \qquad (17)$$

The number of the patterns $M$ enters only the upper limit of the sum in the rhs of Eq. (16). It can be either finite or infinite, but it is not important for the further discussion.

The values $t_\mu^{(k)}$ are arranged in decreasing order. For what follows it is important that the first $k-1$ of these numbers are larger than 1, and the other ones are less than 1:

$$t_1^{(k)} > t_2^{(k)} > ... > t_{k-1}^{(k)} > 1 > t_{k+1}^{(k)} > t_{k+2}^{(k)} > .... \qquad (18)$$

The rhs of Eq. (16) is the result of summing up over the set of functions $\{ f_\mu^{(k)}(y) \}_{\mu \neq k}$. Let us consider the behavior of functions $f_\mu^{(k)}(y)$ (17). It is easy to see that when $y \to \infty$ the denominators $\gamma(\varphi - t_\mu^{(k)})$ of all the functions $f_\mu^{(k)}(y)$ tend to 0. In other words, at the infinity each function $f_\mu^{(k)}(y)$, as well as the rhs of Eq.(16), increases without limit.

The behavior of the function $f_\mu^{(k)}(y)$ for finite values of argument depends on the constant $t_\mu^{(k)}$ in its denominator. If $t_\mu^{(k)} < 1$, then $f_\mu^{(k)}(y)$ is everywhere a continuous function. If $t_\mu^{(k)} > 1$, the function $f_\mu^{(k)}(y)$ has a singular point. In this case for some value $y_\mu^{(k)}$ of the argument of the function $f_\mu^{(k)}(y)$ its denominator is equal to zero:

$$\varphi\left( y_\mu^{(k)} \right) = t_\mu^{(k)} \quad \Leftrightarrow \quad y_\mu^{(k)} = \varphi^{-1}\left( t_\mu^{(k)} \right),$$

where $\varphi^{-1}$ is the inverse of the function $\varphi$. We see that for every $t_\mu^{(k)} > 1$ the function $f_\mu^{(k)}(y)$ has the discontinuity of the second kind in the point $y_\mu^{(k)}$. Since in the series (18) the first $k-1$ numbers $t_\mu^{(k)}$ are greater than 1, it is easy to understand that the rhs of Eq.(16) has the discontinuities of the second kind in $k-1$ points $y_{k-1}^{(k)} < y_{k-2}^{(k)} < ... < y_1^{(k)}$.

For simplicity let us rewrite Eq.(16) using the reciprocal values:

$$\frac{1}{N} = F_k(y), \text{ where } F_k(y) = \left( \sum_{\mu \neq k}^{M} f_\mu^{(k)}(y) \right)^{-1}. \qquad (19)$$

It is evident that nonnegative function $F_k(y)$ in the rhs of Eq.(19) is equal to zero in the points $y_{k-1}^{(k)} < y_{k-2}^{(k)} < ... < y_1^{(k)}$. At the infinity $F_k(y)$ tends to zero. The typical behavior of the function $F_k(y)$ is shown in Fig.7. To the right of the rightmost zero $y_1^{(k)}$, where the inequality $\varphi(y) > t_1^{(k)}$ holds, the function $F_k(y)$ at first increases, and then after reaching its maximum, the function $F_k(y)$ decreases monotonically. Let $y_c^{(k)}$ be the coordinate of the rightmost maximum of the function $F_k(y)$. The value of $F_k(y)$ in the point $y_c^{(k)}$ determines the critical characteristics related to the recognition of the $k$-th pattern. Let us explain what it means.

Generally speaking, Eq.(19) has several solutions. Their number is equal to the number of intersections of the function $F_k(y)$ with the straight line that is parallel to the abscissa axis at the height $1/N$ (see Fig. 7). However, only one of these intersections corresponds to the minimum of the free energy. Namely, this intersection is to the right of the rightmost maximum $y_c^{(k)}$ (see Note in the end of Section II). Other solutions of Eq.(19) have to be omitted.

As an example, in Fig.7 the behavior of the rhs of Eq.(19) for the fifth pattern ($k=5$) is shown for the weights that are the terms of the harmonic sequence $r_\mu = 1/\mu$. Four points $y_4^{(5)} < y_3^{(5)} < y_2^{(5)} < y_1^{(5)}$ are zeros of the function $F_5(y)$. When $y$ is greater than $y_1^{(5)}$, the function $F_5(y)$ at first increases up to the value in the point of the local maximum $y_c^{(5)}$ and then decreases monotonically. The dashed line that is parallel to the abscissa axis is drawn at the height 0.001. When the lhs of Eq.(19) is equal to 0.001, we have $N = 1000$. In other words, for this quasi-Hebbian matrix of the dimensionality $N = 1000$ in the vicinity of the fifth pattern there necessarily is a local minimum. Since the solution of Eq.(19) is large enough, $y_0 \approx 3.5$, the overlap (7) with the pattern is close to 1.

Let us little by little decrease the dimensionality $N$. The dashed straight line will go up, and the solution $y_0(N)$ of Eq.(19) will shift in the region of smaller values. This will go on till $y_0$ coincides with the critical value $y_c^{(5)}$. Just this defines the minimal dimensionality $N_{min}$ for which the local minimum in the vicinity of the fifth pattern still exists. Since for $N < N_{min}$ equation (19) has no solution in the region $y > y_1^{(5)}$, for such $N$ there is no local minimum in the vicinity of the fifth pattern. This means that when $N < N_{min}$ the fifth pattern is not recognized by the network[1].

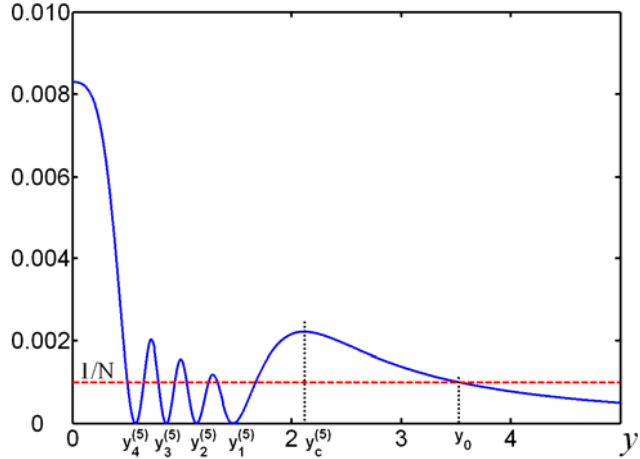

**Fig.7.** The behavior of the function $F_k(y)$ (19) when the weights are equal to $r_\mu = 1/\mu$: $k = 5$, $y_0$ is the solution of the equation (19) for $1/N = 0.001$ and $y_c^{(5)}$ is the critical value of the argument.

Up to now we fixed the number of the pattern $k$ and decreased the dimensionality $N$. It is reasonable to fix the dimensionality $N$ and increase $k$ little by little. We seek its maximal value for which Eq.(19) has a solution. In Fig.8 the behavior of the curves $F_k(y)$ for different values of $k$ is shown. We see that when $k$ increases the critical point $y_c^{(k)}$ shifts to the right, and the value of the maximum of $F_k(y_c^{(k)})$ decreases steadily. It is not difficult to find the maximal value of $k$ for which Eq.(19) has a solution. By $k_m = k_m(N)$ we denote this maximal value of $k$.

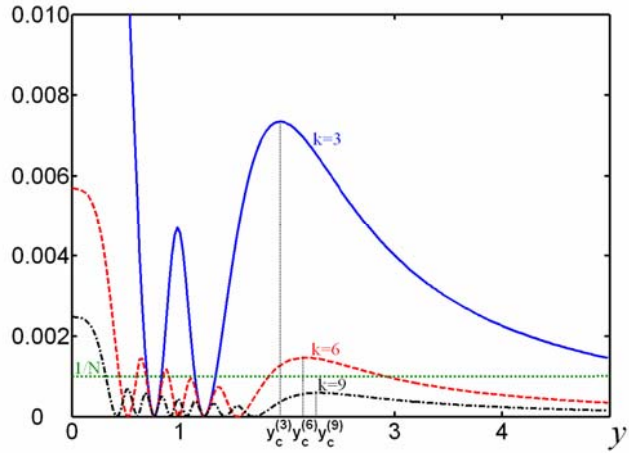

**Fig. 8.** The same as in Fig.7 for $k = 3, 6, 9$. When $k$ increases the critical point $y_c^{(k)}$ shifts to the right, and the value of the maximum $F_k(y_c^{(k)})$ decreases.

For a given dimensionality $N$ the pattern with the number $k_m$ is the last in whose vicinity there is a local minimum. For $k < k_m$ Eq.(19) has a solution to the right of the rightmost maximum $y_c^{(k)}$ as well. Consequently, these patterns would also be recognized. On the contrary, for $k > k_m$ Eq.(19) has no solution in the region $y > y_c^{(k)}$. Consequently, the patterns with such numbers would not be recognized.

---

[1] Let us note that when the dimensionality $N$ decreases, some patterns would be forgotten. This is in agreement with the well-known property of the network: its memory is proportional to the dimensionality of the problem.

Let $r_c$ be the weight corresponding to the pattern with the number $k_m$: $r_c = r_{k_m}$. Our consideration shows that only the patterns, whose weights are not less than the critical value $r_c$ would be recognized. So, the memory of the network is limited, but the catastrophic forgetting does not occur.

This analysis is correct for an arbitrary distribution of the weights $r_\mu$. It provides the algorithm of calculation of the critical value $r_c$. We failed to calculate $r_c$ analytically in the general case. However, in the next Section we examine some specific distributions of the weights for which some analytic results can be obtained.

## V. Some Special Distributions

**1. The weights in the form of geometric sequence.** Let us discuss in detail the case of the weights in the form of decreasing geometric sequence $r_\mu = q^\mu$, where the common ratio $q \in (0,1)$. It was mentioned in [6], [7] that such weights are interesting for applications.

Suppose in Eq.(16) the number of the patterns $M$ tends to infinity: $M \to \infty$. It is natural to assume that in Eq.(16) the first value of the summation index is equal to zero and the first weight is equal to 1: $r_0 = 1$. Then

$$f_\mu^{(k)}(y) \sim \left(\frac{q^\mu}{\varphi_k - q^\mu}\right)^2, \text{ where } \varphi_k = q^k \varphi(y).$$

Now Eq.(16) has the form

$$N = \frac{1}{\gamma^2} \sum_{\substack{\mu=0 \\ \mu \ne k}}^{\infty} \left(\frac{q^\mu}{\varphi_k - q^\mu}\right)^2. \qquad (20)$$

We look for the solution of this equation for large values of the argument when the inequality $\varphi_k = q^k \varphi(y) > 1$ is fulfilled. In the rhs of Eq.(20) we replace summation by integration, and as a result we obtain

$$\lim_{M \to \infty} \sum_{\mu=0}^{M} \frac{q^{2\mu}}{(\varphi_k - q^\mu)^2} = \frac{1}{|\ln q|} \left[\ln\left(\frac{\varphi_k - 1}{\varphi_k}\right) + \frac{1}{\varphi_k - 1}\right].$$

If by $\Phi_k(y)$ we denote the r.h.s of this expression and subtract the term with $\mu = k$, the rhs of Eq. (20) takes the form:

$$\sum_{\mu=0, \mu \ne k}^{\infty} f_\mu^{(k)}(y) = \frac{1}{\gamma^2}\left[\Phi_k(y) - \frac{1}{(\varphi-1)^2}\right].$$

It is convenient to pass to the reciprocals functions in the both sides of Eq. (20). Then we obtain an analogue of Eq. (19):

$$\frac{1}{N} = \frac{\gamma^2 (\varphi-1)^2}{(\varphi-1)^2 \cdot \Phi_k - 1}. \qquad (21)$$

By solving Eq.(21) for a given $q$ with the aid of computer simulations, one can find the number of the last pattern that can be recognized. We denote this number as $k_m(N,q)$. We are interested in the value of $q$ that defines the maximal value of $k_m(N,q)$. Let $q_m$ denote this optimal value, and let $k_m(N)$ be the corresponding number of the patterns: $k_m(N) = k_m(N, q_m) = \max_q k_m(N,q)$. It is evident that such an optimal value $q_m$ has to exist. Indeed, as long as the number $q$ is small, the number of the recognized patterns is also small. In this case the weights $r_\mu = q^\mu$ decrease very quickly when $\mu$ increases, and only the very first patterns will be recognized. It is even possible that it will be the first pattern only. On the contrary, when $q \to 1$, our model becomes very close to the standard Hopfield model. For sufficiently large values of $q \approx 1$ no one of the patterns is recognized due to the catastrophic forgetting. Consequently, there must be the optimal value $q_m \in (0,1)$, so that the number of recognized patterns would be maximal: $q_m = q_m(N)$.

It is easy to find out the critical value $q_c$ for which only the first pattern would be recognized. It may be shown that the following estimate for $q_c$ is true:

$$q_c = 1 - \delta, \text{ where } \delta \approx \frac{1}{0.329N}.$$

For $q > q_c$, the network ceases to recognize patterns at all.

Up to now we did not succeed in an analytical calculation of $q_m(N)$. However, this estimate can be obtained by solving Eq.(21) numerically for different values of $q$ and $N$. In the left panel of Fig.9 the dependence of the ratio $k_m(N,q)/N$ on $q$ for three dimensionalities $N$ is shown. We see that the curves have distinct points of maximum, but the value of all the maximums are approximately the same:

$$\lim_{N\to\infty} k_m(N)/N \approx 0.05. \tag{22}$$

It can be shown that for the optimal value $q_m$ the following estimate is valid

$$q_m \approx 1 - 2.75 \cdot \delta. \tag{23}$$

The expression (22) shows that the maximal number of patterns that can be stored by the network is $M_c \approx 0.05 \cdot N$. It is less than the storage capacity of the standard Hopfield model ($0.138 \cdot N$), but the catastrophic destruction of the memory does not occur.

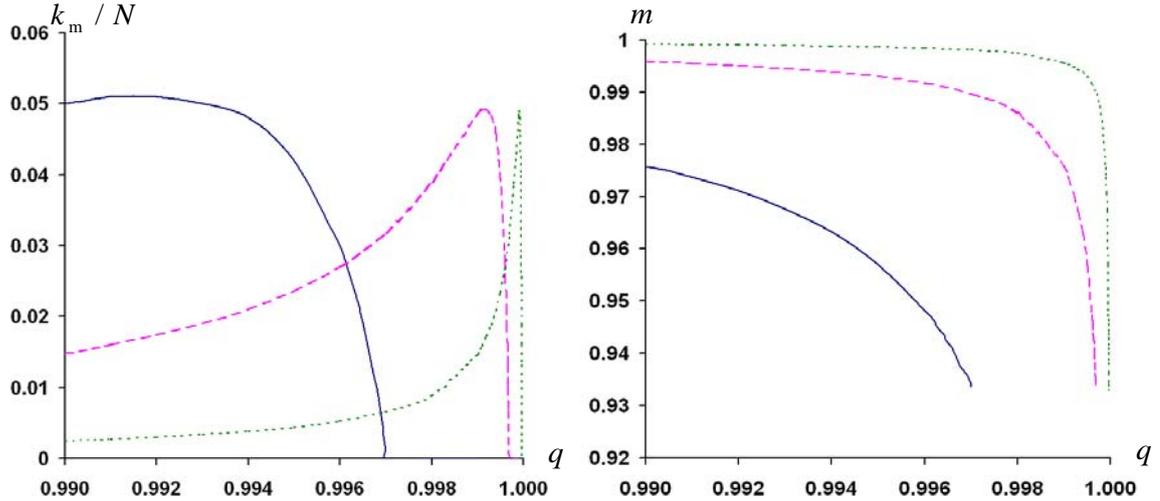

**Fig.9.** For three dimensionalities $N$ we show: the dependence $k_m/N$ on $q$ (left panel); the corresponding values of the overlap for the last recognized pattern (right panel). In both panels solid lines correspond to the dimensionality $N=1000$, the dashed lines correspond to $N=10000$, the point lines correspond to $N=100000$.

The right panel of Fig.9 shows the dependence of the overlap on $q$ for the last recognized pattern (whose number is $k_m$). In the point of the solution "breakdown" all overlaps have approximately the same values $m_c \approx 0.933$.

We can move a step forward in the analytical calculations if for $q \to 1$ and large values of $y$ (let us say, for $y \geq 2$) we use the asymptotic expression for the rhs of Eq.(21):

$$\frac{1}{N} = g^2 \frac{2q^{2k} |\ln q|}{1 - 2q^{2k} |\ln q|}, \text{ where } g = \frac{erf\ y}{\sqrt{2}y} \approx \frac{1}{\sqrt{2}y}. \tag{24}$$

In Fig.10 the right-hand sides of Eqs. (21) and (24) are shown for different values of $k$ and $q$. We can see that for large values of the argument ($y \geq 2$) the rhs of Eqs. (21) and (24) practically coincide.

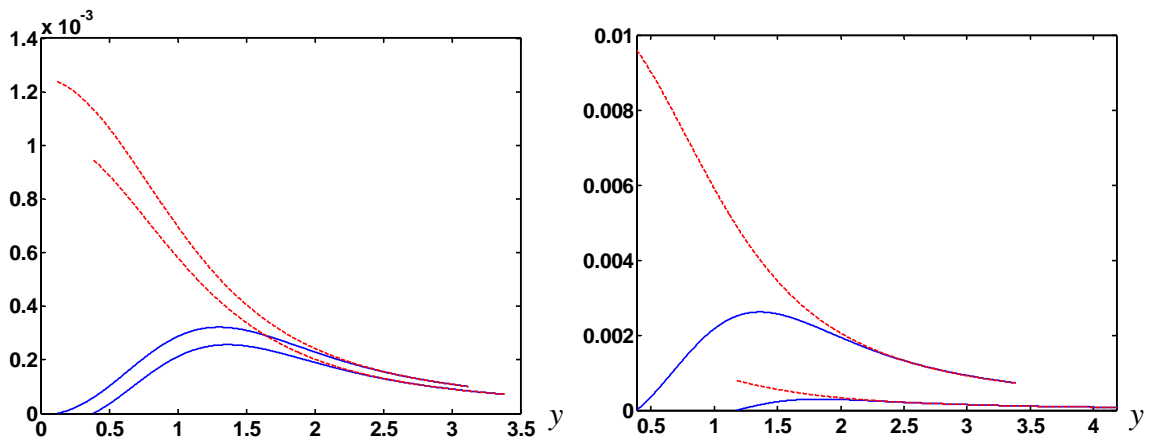

**Fig. 10.** Rhs of Eq.(21) (solid lines) and rhs of Eq.(24) (dashed lines) for different values of $k$ and $q$:

in the left panel $q = 0.999$, in the right panel $q = 0.99$. In both panels the lower curves correspond to $k = 10$, and the upper ones to $k = 100$.

Equation (24) allows us to write down the explicit expression for $k$:

$$k = \frac{\ln[2(Ng^2+1)|\ln q|]}{2|\ln q|}. \quad (25)$$

We can calculate the maximal value of $k(q)$. When $q_0 \approx 1 - \frac{e}{2(Ng^2+1)}$, the maximal value of $k$ is equal to $k_0 \approx Ng^2/e \approx N/2ey^2$. Now we require the conditions of the perfect recognition to be fulfilled. That means that the difference between the pattern and the nearest local minimum has to be less than 1: $m = erf(y) > 1 - 1/N$. We obtain that in this case $y \approx \sqrt{\ln(N/4)}$. Substituting this value of $y$ into the expression for $k_0$ and denoting by $k_p$ the maximal number of perfectly recognized patterns, we have:

$$\lim_{N\to\infty} k_p/N = \lim_{N\to\infty} 0.5e^{-1}/\ln(N/4) = 0. \quad (26)$$

Thus, the requirement of the perfect recognition decreases the storage capacity of the network substantially: compare expressions (26) and (22). The same is true for the standard Hopfield model.

In Fig.11 we plot three graphs corresponding to the dimensionality $N = 1000$. The solid line shows the dependence of the value of the recognized patterns $k_m$ on $q$. The dashed line shows the dependence of the value of the perfectly recognized patterns $k_p$ on $q$. This dependence is calculated by substituting the expression $y \approx \sqrt{\ln(N/4)}$ into equation (25). Markers in Fig.11 show the numerical results for the number of perfectly recognized patterns. The experimental results are averaged over 500 random matrices. On the whole the agreement with the theory is quite good.

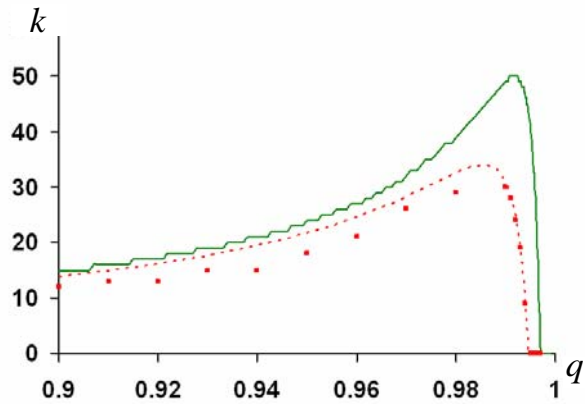

**Fig. 11.** For $N = 1000$ we show the dependencies on $q$ of the maximal number of the recognized patterns $k_m$ (solid line), the number of the perfectly recognized patterns $k_p$ (dotted line). Markers are the experimental numbers of the perfectly recognized patterns.

**2. The weights in the form of harmonic series:** $r_\mu = 1/\mu$. In this case we also can let the upper limit in the sum in the rhs of Eq.(16) tend to infinity: $M \to \infty$. Next, the function $f_\mu^{(k)}$ (17) has the form:

$$f_\mu^{(k)}(y) \sim \left(\frac{k}{\mu\varphi - k}\right)^2 = \frac{a_k^2}{(\mu - a_k)^2},$$

where $a_k \equiv a_k(y) = k/\varphi(y) < 1$. Then

$$\sum_{\mu \neq k}^{\infty} f_\mu^{(k)}(y) = \frac{a_k^2}{\gamma^2} \sum_{\mu \neq k}^{\infty} \frac{1}{(\mu - a_k)^2}. \quad (27)$$

The sum in the rhs of Eq.(27) resembles the Hurvitz zeta function $\zeta(s,a)$ for s=2 (see [13]):

$$\zeta(s,a) = \sum_{\mu=0}^{\infty} \frac{1}{(\mu+a)^s}.$$

Usually one examines zeta function as a function of positive argument $a$, but in Eq.(27) this argument is negative. It is not difficult to show that for $a \in (0,1)$ the equality $\zeta(2,-a) = \dfrac{1}{a^2} + \zeta(2, 1-a)$ is true. Now after simple transformations we obtain equation analogous to Eq.(19):

$$\frac{1}{N} = \frac{\gamma^2(\varphi-1)^2}{a_k^2(\varphi-1)^2 \zeta(2, 1-a_k) - 1}. \tag{28}$$

As before, by $F_k(y)$ we denote the rhs of Eq.(28). As an example, in Fig.7 we show the graph of the function $F_5(y)$. We are interested in the behavior of the function $F_k(y)$ in the region of large values of its argument, where the inequality $a_k(y) < 1$ is fulfilled. This inequality means that $\varphi(y) > k$. Moreover, we have to examine the values of $y$ that are to the right of the rightmost maximum $F_k(y)$ only.

By solving Eq. (28) numerically for each $N$ we can find the maximal number of the pattern $k_m(N)$, which is still recognized by the network. Patterns whose numbers are less than $k_m(N)$ are also recognized, but patterns whose weights are larger than $k_m(N)$ are not recognized. Here the critical value of the weight $r_c$ is equal to the reciprocal of $k_m(N)$: $r_c = 1/k_m(N)$. On the other hand, in Ref. [14], [15] the critical value of $r_c$ was estimated as $r_c = \sqrt{2\ln N / N}$. The last result was obtained with the aid of probability-theoretical technique known as "signal-to-noise ratio" under the most general assumptions about the weights $\{r_\mu\}_{\mu=1}^\infty$. After adjusting the normalization conditions we obtain that $k_m(N)$ has to be equal to $k_c(N)$ ($k_c(N)$ is obtained from the estimate $r_c = \sqrt{2\ln N / N}$):

$$k_c(N) = \frac{1}{\pi}\sqrt{\frac{3N}{\ln N}}. \tag{29}$$

In Fig. 12 we show the graphs $k_m(N)$ (solid line) and $k_c(N)$ (dashed line) for different values of $N$. We see that in the wide range of values of $N$ the curves are close to each other: $k_m(N) \approx k_c(N)$.

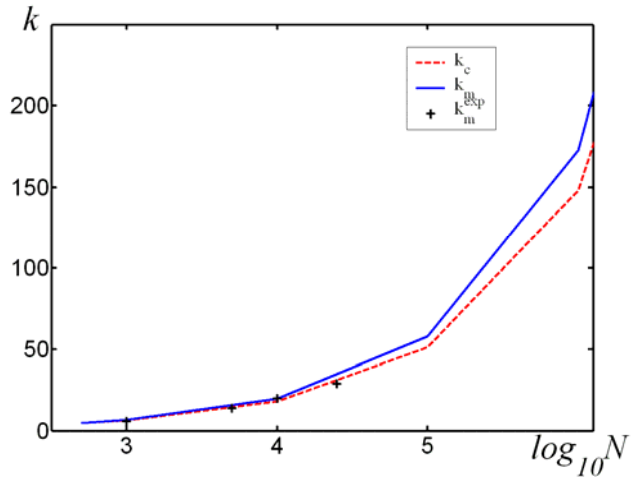

**Fig.12.** The dependence of $k_m$ (solid line) and $k_c$ (dashed line) on $N$. Markers show experimental values of $k_m^{\exp}$.

In Fig.12 markers indicate the experimental values of the critical numbers of the patterns for the dimensionalities $N = 1000, 5000, 10000$ and $25000$. For each given $N$ we generated a random matrix with the weights in the form of the terms of the harmonic series and determined the maximal number of the pattern, which was a fixed point. As a result of averaging over 10 random matrices we obtained the value of $k_m^{\exp}$, which is an experimental estimate for $k_m$. We see that our experiment is in a good agreement with the theoretical prediction.

Thus, the storage capacity of such a network tends to zero:

$$\lim_{N \to \infty} k_m(N)/N \sim \lim_{N \to \infty} 1/\sqrt{N \ln N} = 0. \tag{30}$$

It is not surprising that the storage capacity of this network is much less than the storage capacity of a network with the weights in the form of the terms of the geometric sequence. The point is that the harmonic series is dominated by the geometric sequence. In Fig. 13 for different values of $N$ we show the difference between the terms of the harmonic series $1/k$ and the geometric sequence with $q = q_m(N)$ (23). For convenience along the abscises axis we put the values of $k/N$. We see that for $k < N/2$ the terms of the geometric sequence are noticeably larger than the terms of the harmonic series. Only beginning from $k \approx N/2$ the terms of these two sequences become almost equal. In other words, an overwhelming number of the weights in the form of the terms of the harmonic series are substantially less than the weights in the form $q_m^k$.

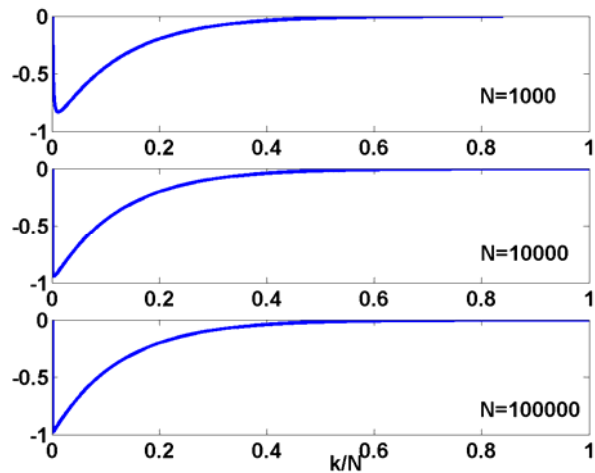

**Fig.13.** The differences $1/k - q^{k-1}$ when $k \in [1, N]$, and $q$ is equal to the optimal value $q_m(N)$ (23). The values of $N$ are shown in the panels. Along the abscises axis is the value of $k/N$.

**3. The weights in the form of arithmetic progression:** $r_\mu = 1 - (\mu - 1)d$, $\mu = 1,...,M$. The weights in the form of an arithmetic progression are interesting because they can serve as an approximation of the case of random weights distributed uniformly inside the interval (0,1). Since the weights have to be positive, the upper limit of the sum in the rhs of Eq. (16) has to be finite: $M < \infty$. The parameters $M$ and $N$ are independent: the number of the patterns $M$ can be more or less than the dimensionality $N$. Another free parameter is the common difference $d$. The maximal possible common difference $d_m = 1/M$ is defined from the condition that the smallest weight $r_M$ would be positive. At first we analyze the case $d = d_m$, and after that we examine the case $d = d_m/g$, where $g > 1$.

Generally speaking, the weights in the form of the arithmetic progression are larger than the weights in the form of the geometric sequence. This is clearly seen from Fig.14. Here for different dimensionalities $N$ we show the behavior of the difference $1 - (k-1)d - q^{k-1}$, when $d = 1/N$, the common ratio is $q = q_m(N)$ (23), and $k$ varies from 1 to $N$. We see that everywhere the terms of the arithmetic progression exceed the terms of the geometric sequence. One can expect that the storage capacity of a network with the weights in the form of the terms of the arithmetic progression exceeds the capacity of the network with the weights in the form of the geometric sequence. In what follows we will see that it is indeed the case.

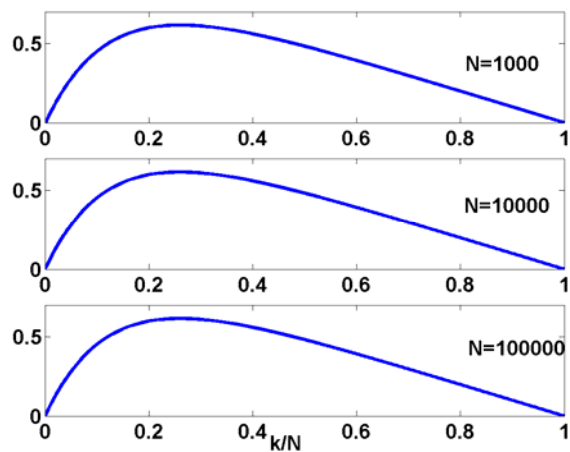

**Fig.14.** The difference of the weights $1 - (k-1)d - q^{k-1}$ for $d = 1/N$, $q = q_m(N)$ (23), $k \in [1, N]$. Along the abscissa axis is the value of $k/N$.

Let us transform the rhs of the equation (16) in the same way as it was done in subsections 1 and 2. Assuming that $M$ is very large we replace summation by integration. Then the rhs of this equation is

$$\sum_{\mu \neq k}^{M} f_{\mu}^{(k)}(y) = \frac{1}{\gamma^2}\left[ M \cdot \Phi_k(y) - \frac{1}{(\varphi-1)^2} \right],$$

where

$$\Phi_k = \frac{2\varphi_k - 1}{\varphi_k - 1} + 2\varphi_k \ln\left(\frac{\varphi_k - 1}{\varphi_k}\right), \quad \varphi_k = \varphi(y)\left(1 - \frac{k}{M}\right).$$

If we introduce a relative number of the pattern $\kappa = k/M$ we obtain the equation that is analogous to Eqs.(21) and (28):

$$\alpha = \frac{\gamma^2}{\dfrac{2\varphi_k - 1}{\varphi_k - 1} + 2\varphi_k \ln\left(\dfrac{\varphi_k - 1}{\varphi_k}\right)}, \tag{31}$$

where $\varphi_k = \varphi(y)(1-\kappa)$. Here $\alpha = M/N$ is the standard notation for the load parameter. We solve this equation in the region of the large values of $y$ where $\varphi_k = \varphi(y)(1-\kappa) > 1$. We are interested in the solutions which are to the right of the point, where the expression in the rhs of Eq.(31) is maximal.

In Fig.15 we show the behavior of the rhs of Eq.(31) for three different values of $\kappa = 0, 0.49$ and $0.6$. The solid line corresponds to $\kappa = 0$. It characterizes the conditions of recognition of the very first pattern with the largest weight $r_1 = 1$. We see that the maximal value of the load parameter $\alpha$ for which the first pattern is still recognized is $\alpha_c(0) \approx 0.47$. No pattern will be recognized, if the load parameter of the network is larger that this value, i.e. when $\alpha > \alpha_c(0)$. The dashed line corresponds to $\kappa = k/M = 0.49$. To recognize 49% of the patterns written down into the connection matrix the load parameter $\alpha$ has to be less than 10%: $\alpha_c(0.49) \approx 0.09$. To recognize 60% of the patterns the load parameter has to be even smaller (see the chain line): $\alpha_c(0.6) \approx 0.05$. When $\kappa$ increases the critical value of the load parameter $\alpha_c(\kappa)$ decreases monotonically. This dependence can be easily understood: the greater the part of the patterns $\kappa$ that have to be recognized, the greater must be the dimensionality of the problem $N$, and, consequently, the less must be the value of the load parameter $\alpha$.

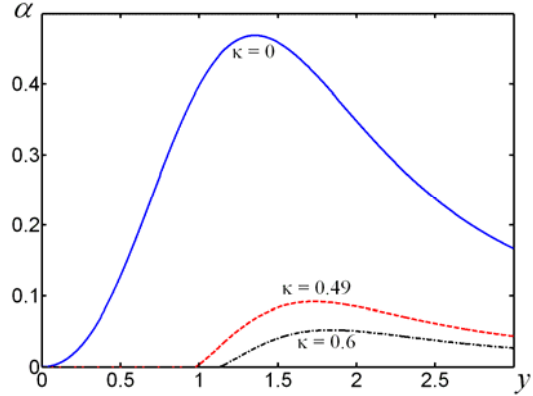

**Fig. 15.** The rhs of equation (31) for different values of $\kappa$: $\kappa = 0$ (solid line), $\kappa = 0.49$ (dashed line) and $\kappa = 0.6$ (chain line). The maximum on the curve defines the critical value of the load parameter $\alpha_c(\kappa)$.

As the storage capacity of the network one usually understands the ratio of the recognized patterns $k$ to the dimensionality of the problem $N$: $k/N = \kappa \cdot \alpha$. In Fig. 16 we show the dependences of the functions $\alpha_c(\kappa)$ and $\kappa \cdot \alpha_c(\kappa)$ on $\kappa$, where by $\alpha_c(\kappa)$ we define the maximal value of the function in the rhs of Eq.(31) for the given $\kappa$. We see that when $\kappa$ increases the function $\alpha_c(\kappa)$ decreases monotonically (the lower panel in Fig. 16). In the same time the curve $k/N = \kappa \cdot \alpha_c(\kappa)$ has a pronounced maximum when $\kappa_m \approx 0.3$. The value of this maximum is

$$k_m/N \approx 0.06. \tag{32}$$

The expression (32) defines the optimal value of the storage capacity of a network whose weights are the terms of the arithmetic progression. Comparing (32) with the analogous expression (22), we see that if the weights are in the form of the terms of the arithmetic progression, the larger storage capacity can be obtained.

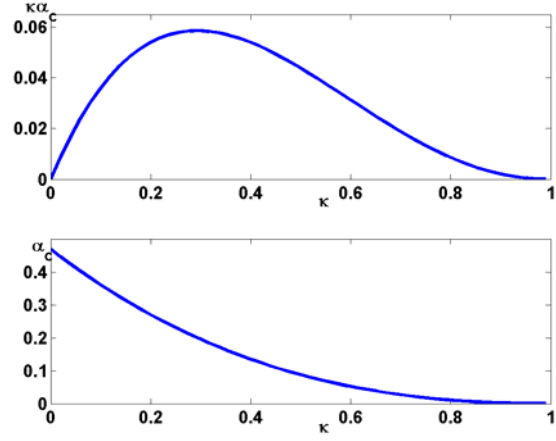

**Fig. 16.** The dependence of the critical value of the load parameter $\alpha_c(\kappa)$ and the storage capacity $k/N = \kappa\alpha_c(\kappa)$ on $\kappa$. The storage capacity reaches its maximum $k_m/N \approx 0.06$, when $\kappa \approx 0.3$.

In the abovementioned case the common difference is equal to $d_m = 1/M$ and the weights decrease up to a very small value $r_M = 1/M$. Let us reduce the common difference. Suppose $d = d_m/g$, and $g > 1$. Then the weights are distributed in the interval $(b,1]$, where $b = 1 - 1/g > 0$. It is easy to verify that now Eq.(32) takes the form:

$$\alpha = \frac{\gamma^2}{1 + \dfrac{\varphi_k^2}{(\varphi_k - 1)(\varphi_k - b)} + \dfrac{2\varphi_k}{1-b}\ln\left(\dfrac{\varphi_k - 1}{\varphi_k - b}\right)}, \qquad (33)$$

where $\varphi_k = \varphi(y)(1 - \kappa/g)$. When $g = 1$, Eq.(33) transforms into Eq.(31).

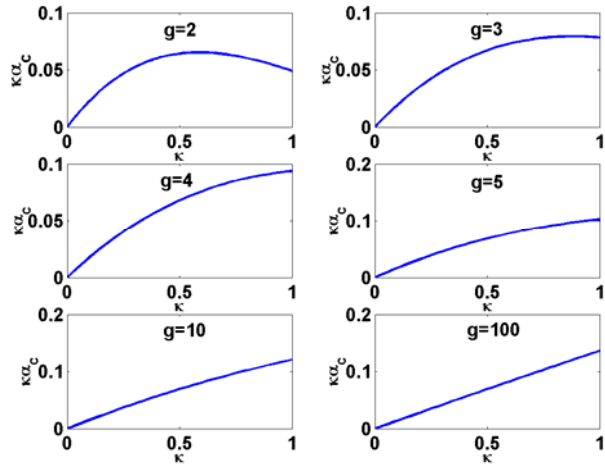

**Fig. 17.** The behavior of the storage capacity $k/N = \kappa\alpha_c(\kappa)$ for different values of the parameter $g$.

In Fig. 17 we show how the curve $\kappa\alpha_c(\kappa)$ depends on the parameter $g$ (in the upper panel of Fig.16 is the same curve for $g = 1$). The maximum of this curve corresponds to the maximal storage capacity of the network $k_m/N$. We see that when $g$ increases the value of the maximum also increases, and the point of the maximum shifts to the right toward the value $\kappa = 1$. Beginning from $g \approx 4$ the maximum of the curve always is in the point $\kappa = 1$. Further increase of $g$ only leads to the rectification of the curve. Little by little it transforms into a straight line, and the maximal storage capacity tends asymptotically to the value of the storage capacity of the standard Hopfield model: $\alpha_c \approx 0.138$. The larger $g$ the less the common difference $d$, and also the less the difference between the maximal weight $r_1 = 1$ and the minimal weight $r_M = 1 - 1/g$. This means that our model more and more resembles the standard Hopfield model.

## Conclusions

In the standard Hopfield model a stage of learning of the network and a working stage are absolutely separated. At the learning stage the Hebb connection matrix is constructed with the aid of input patterns. At the working stage the network recognizes the distorted versions of the input patterns. Before adding the new patterns in the connection matrix one must verify that the new number of the patterns does not exceed the critical value $M_c = 0.138 \cdot N$. Otherwise the memory of network will be destructed.

Usually in nature learning processes are continuous ones. An artificial memory also has to work even if it obtains new information continuously. For this reason the catastrophic destruction of the memory of the Hopfield model due to its overfilling is absolutely inadmissible.

The method of eliminating of the catastrophic destruction, which we propose in our paper, seems to be very attractive: it allows the network to learn even during the working stage. Each input pattern modifies the matrix elements according to the standard Hebbian rule. If the input pattern is the same as the one written down previously, its weight increases by 1. If the pattern is new, it is written down into the connection matrix with the initial weight equals to 1. The current connection matrix is modified uninterruptedly. Since the memory of the network consists of the patterns whose weights are larger than the critical value $r_c$, one should not worry about its overfilling. This critical $r_c$ depends on the current weights distribution.

Let the weight of a pattern be less than $r_c$, but this pattern has to be recognized by the network. For this purpose it is sufficient to increase the weight of this pattern making it larger than the critical value $r_c$. It is possible that at the same time some other patterns cease to be recognized. Such replacement of patterns by other ones does not contradict to the common sense. It corresponds to the general conception of the human memory.

## Acknowledgments

The work was supported by the program of the Presidium of the Russian Academy of Sciences (project 2.15) and in part by the Russian Basic Research Foundation (grant 12-07-00295). The authors are grateful to Dr. Inna Kaganova for stimulating discussions and for the help during the course of the work.